# The structural, elastic and optical properties of ScM (M = Rh, Cu, Ag, Hg) intermetallic compounds under pressure by ab initio simulations

Md. Lokman Ali[1], Md. Zahidur Rahaman[2], Md. Atikur Rahman[3]

*[1, 2, 3]Department of Physics, Pabna University of Science and Technology, Pabna-6600, Bangladesh*

## Abstract

The influence of pressure on the structural and elastic properties of ScM (M = Rh, Cu, Ag, Hg) compounds has been performed by using ab initio approach pseudopotential plane-wave method based on the density functional theory within the generalized gradient approximation (GGA). The optical properties have been investigated under zero pressure. It is found that the optimized lattice parameters for all metals are in good agreement with the experimental data and other available theoretical values. We obtained three independent elastic constants $C_{ij}$ ($C_{11}$, $C_{12}$ and $C_{44}$) and various elastic parameters such as bulk modulus $B$, shear modulus $G$, Young's modulus $E$, $B/G$, Poisson's ratio $v$ and anisotropy factor $A$ as a function of pressure. In addition, the mechanical stability and ductile/brittle nature are also investigated from the calculated elastic constants. The study of optical properties reveals that all of these compounds possess good absorption coefficient in the high energy region and the refractive index of all these compounds is higher in the low energy region and gradually decreased in the high energy region. All these calculations have been carried out using the CASTEP computer code.

## 1. Introduction

Scandium-based intermetallic compounds have significant interest among the research community due to their structural and remarkable physical properties. The Sc-based compounds are of technological interest due to their prominent applications in the production of material for spacecraft and electronic devices [1]. Scandium alloy is one of the lightest materials than most other metals. It is also resistant to corrosion and has a high melting point. Scandium and its alloys have been used for a variety of application such as aerospace industry, baseball bats, lacrosse sticks and bicycle frames. Furthermore, these compounds also possess many other attractive properties such as high ductility, high tensile strength and high fracture toughness at room temperature [2]. A number of theoretical and experimental works have been performed on structural, electronic and phonon properties of Sc-TM (TM = Ag, Cu, Pd, Rh, Ru) compounds [3-8]. Bushra Fatema et al. [9] have been performed a density functional study on structural, elastic, electronic and mechanical properties of Sc-TM (TM = Co, Rh, Ir, Ni, Pd, Pt, Zn, Cd and Hg) compounds under zero pressure using WIEN2K code [10]. The variation of elastic moduli with different hydrostatic pressure in the B2 phase for XRh (X = Sc, Y, Ti and Zr) compounds have been studied by Bushra Fatima et al. [11] using the FP-LAPW method.

However, the above mentioned all the studies have been done at zero pressure. To the best of our knowledge the optical properties of ScM (M = Rh, Cu, Ag and Hg) compounds are still unexplored. Furthermore, the pressure effects on structural and elastic properties of ScM (M = Rh, Cu, Ag and Hg) compounds have not been reported up to now. It is well known that, pressure plays a vital role in affecting the physical properties of solids such as structural and elastic parameters [12-14].

---

[3]Corresponding Author: atik0707phy@gmail.com

J. Kang et al. [15] and H.Y. Xiao et al. [16] reported that high pressure contributes to the transition of phase and changes in physical properties of material. So the study of pressure effects on ScM is important and significant. This work will help to run further study on the physical properties of these materials under pressure, and also provide reference data for future experimental work.

In our present work, we have investigated the pressure effects on the structural and elastic properties of the ScM intermetallic compounds for the first time. In addition, the optical properties under zero pressure are also calculated and discussed. The organization of the paper is as follows: Section 2 presents the computational details. Section 3 presents the result and discussion. Finally, the summary of our results is given in section 4.

## 2. Computational methods and details

The Cambridge serial total energy package (CASTEP) [17], a first principles calculations plane wave pseudo-potentials method based on density function theory (DFT) [18] is used to investigate the behaviour of ScM (M = Rh, Cu, Ag and Hg) compounds under hydrostatic pressures. The effects of exchange-correlation energy function were considered by the generalized gradient approximation (GGA) [20] of the Perdew-Burke-Ernzerhof (PBE) form [21]. The plane wave basis set with an energy cut-off of 340 eV for all cases is used. For the sampling of the Brillouin zone, the Monkhorst-pack [22] k-points of ($8 \times 8 \times 8$) are used. For smooth convergence the parameters are set as follows: (i) $1.0 \times 10^{-5}$ e/atom for the total energy, (ii) 0.03 eV/Å for the maximum force, (iii) 0.05 GPa for the maximum stress and (iv) the maximum ionic displacement within 0.001 Å.

The elastic stiffness constants of cubic ScM (M = Rh, Cu, Ag, Hg) compounds were determined by the stress-strain method [23] with a finite value. The convergence tolerances were set as follows: (i) $2.0 \times 10^{-6}$ eV/atom for energy (ii) maximum ionic force within 0.006 eV/Å, (iii) maximum ionic displacement within $2.0 \times 10^{-4}$ Å. The maximum strain amplitude is set to be 0.003 in the present all calculations. The values were found to be carefully tested and sufficient for studying the structural, elastic and optical properties of SCM (M = Rh, Cu, Ag and Hg) intermetallic compounds.

## 3. Results and Discussion

### 3.1. Structural properties

The ScM (M = Rh, Cu, Ag, Hg) intermetallic group possess cubic lattice of CsCl type structure which has space group Pm-3m (221). The equilibrium lattice parameter has a value of 3.20, 2.25, 3.41 and 3.48 Å for ScRh, ScCu, ScAg and ScHg respectively [24, 25, 25, and 27]. The atomic positions and lattice constants of these compounds had been optimized as a function of normal stress by minimizing the total energy. The optimized structure of ScM (M = Rh, Cu, Ag, Hg) is shown in Fig.1. The Calculated structural parameters of ScRh, ScCu, ScAg and ScHg at P = 0 GPa are given in Table 1 along with the available experimental and other theoretical values. It is evident from Table 1 that our present calculated values are very close to both experimental and other theoretical values. The calculated lattice constants of our present work are 3.23, 3.28, 3.48 and 3.57 Å for ScRh, ScCu, ScAg and ScHg respectively which exhibits 0.92, 0.91, 2.01 and 2.5 % deviation from the experimentally measured lattice parameters as shown in Table 1. The different calculation method and the different condition can be the reason for this existing discrepancy. However our calculated values match well with the experimental and other theoretical values which indicate the reliability of our present DFT based calculations.

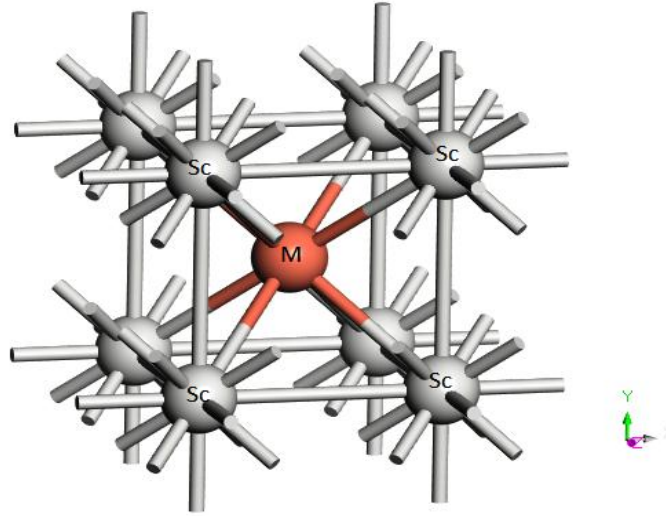

**Fig. 1.** The crystal structure of ScM (M = Rh, Cu, Ag, Hg).

**Table 1.** The calculated equilibrium Lattice constant "$a_0$", bulk modulus "$B$" and its first pressure derivative "$dB/dP$" and unit cell volume "$V_0$" of ScM (M = Rh, Cu, Ag, Hg) at zero pressure.

| Materials | Referance | $a_0$ (Å) | $B(GPa)$ | $dB/dP$ | $V_0$ (Å$^3$) | Deviation from Expt. (%) |
|---|---|---|---|---|---|---|
| **ScRh** | Present | 3.233 | 134.57 | 4.11 | 33.80 | 0.92 |
| | Expt. [24] | 3.200 | 157.30 | -- | -- | |
| | Theory [2] | 3.220 | 161.00 | 4.01 | -- | |
| **ScCu** | Present | 3.289 | 98.28 | 3.23 | 35.58 | 0.91 |
| | Expt. [25] | 3.257 | -- | -- | -- | |
| | Theory [2] | 3.279 | 84.70 | 3.91 | -- | |
| **ScAg** | Present | 3.488 | 65.53 | 4.39 | 42.5 | 2.01 |
| | Expt. [25] | 3.414 | -- | | -- | |
| | Theory[26] | 3.439 | 99.00 | 4.40 | -- | |
| **ScHg** | Present | 3.576 | 79.65 | 3.58 | 45.75 | 2.5 |
| | Expt. [27] | 3.480 | -- | | -- | |
| | Theory[9] | 3.520 | 79.10 | 4.80 | -- | |

For investigating the influence of external pressure on the crystal structure of ScM (M = Rh, Cu, Ag, Hg), we have studied the variations of the lattice parameters and unit cell volume of ScM (M = Rh, Cu, Ag, Hg) with different pressure. The calculated structural parameters of ScM (M = Rh, Cu, Ag, Hg) at different hydrostatic pressures is shown in Table 2. Fig. 2(a) illustrates the variations of lattice parameters of ScM (M = Rh, Cu, Ag, Hg) with pressure. It is evident that the ratio $a/a_0$ decreases with the increase of pressure, where $a_0$ is the equilibrium lattice parameter at zero pressure. However, with the increase of pressure, the distance between atoms is reduced. As a result the repulsive interaction between atoms is strengthened, which leads to the difficulty of compression of the crystal under pressure. The pressure-volume curves of ScM (M = Rh, Cu, Ag, Hg) intermetallic group are plotted in Fig. 2(b). From these curve we observe that with the increase of pressure volume of ScM (M = Rh, Cu, Ag, Hg) intermetallics are decreased gradually. The obtained pressure-volume data are fitted to a third-order Birch-Murnaghan equation of state (EOS) [28].

**Table 2.** The calculated Lattice constant "*a*", and unit cell volume "*V*" of ScM (M = Rh, Cu, Ag, Hg) at different pressure.

| Materials | *P* (GPa) | *a* (Å) | *V* (Å$^3$) |
|---|---|---|---|
| **ScRh** | 20 | 3.206 | 32.95 |
|  | 40 | 3.042 | 28.17 |
|  | 60 | 2.980 | 26.48 |
| **ScCu** | 20 | 3.090 | 29.53 |
|  | 40 | 2.987 | 26.65 |
|  | 60 | 2.911 | 24.67 |
| **ScAg** | 10 | 3.339 | 37.23 |
|  | 20 | 3.259 | 34.63 |
| **ScHg** | 20 | 3.340 | 37.26 |
|  | 40 | 3.226 | 33.59 |
|  | 60 | 3.148 | 31.19 |

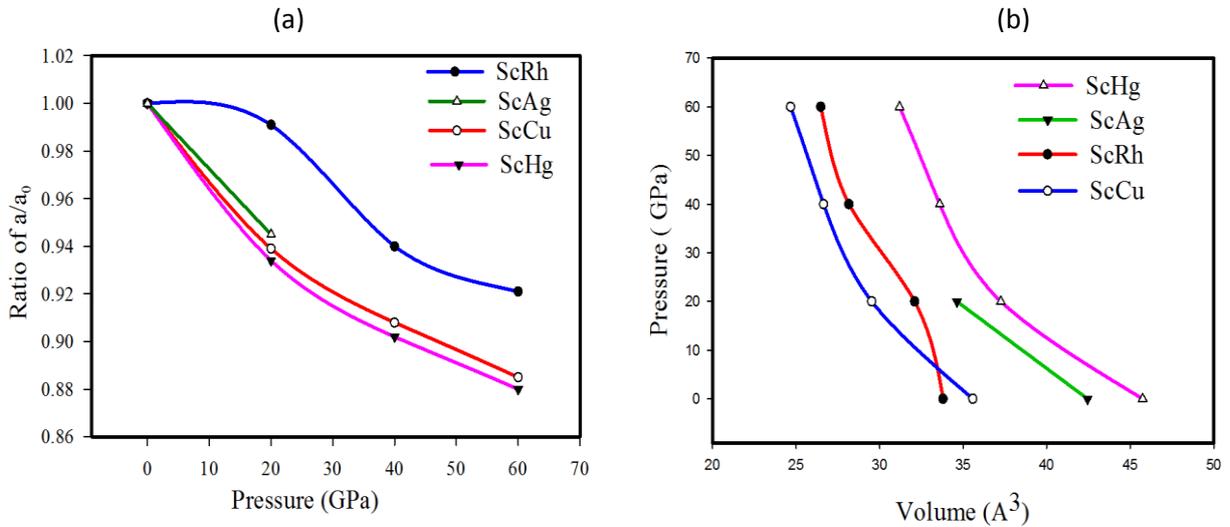

**Fig. 2.** Variation of lattice parameters as a function of pressure (a). Birch-Murnaghan equation of state for ScM (M = Rh, Cu, Ag, Hg) (b).

*3.2. Elastic properties*

Elastic properties play a vital role to describe the nature of forces in solids [29]. These properties provide useful information concerning the structural stabilities, bonding characteristics, anisotropic character of binding and the stiffness of crystals. Q. Chen et al. [30] reported that for a cubic crystal have three independent elastic constants such as, $C_{11}$, $C_{12}$ and $C_{44}$. The calculated elastic constants at zero pressure are presented in Table 3.

**Table 3.** Calculated elastic constants $C_{ij}$ (GPa) and elastic parameters (B, G, E, B/G, ν and A) of (a) ScRh (b) ScCu (c) ScAg (d) ScHg at P = 0 GPa.

| Materials | Referance | $C_{11}$ | $C_{12}$ | $C_{44}$ | B | G | E | B/G | ν | A |
|---|---|---|---|---|---|---|---|---|---|---|
| **ScRh** | Present | 284.36 | 59.68 | 48.48 | 134.57 | 68.38 | 175.44 | 1.96 | 0.56 | 0.43 |
| | Expt. [24] | 242.00 | 114.9 | 54.00 | -- | -- | -- | -- | -- | -- |
| | Theory[24] | 232.40 | 108.3 | 51.8 | 141.52 | 40.48 | 110.88 | 3.49 | 0.37 | 0.83 |
| **ScCu** | Present | 118.96 | 56.04 | 49.11 | 77.01 | 41.07 | 104.61 | 1.87 | 0.27 | 1.56 |
| | Expt. | -- | -- | -- | -- | -- | -- | -- | -- | -- |
| | Theory [2] | -- | -- | -- | 84.7 | -- | -- | -- | -- | -- |
| **ScAg** | Present | 86.58 | 55.01 | 32.10 | 65.53 | 24.14 | 64.49 | 2.71 | 0.33 | 2.03 |
| | Expt. | -- | -- | -- | -- | -- | -- | -- | -- | -- |
| | Theory [2] | -- | -- | -- | 79.3 | -- | -- | -- | -- | -- |
| **ScHg** | Present | 84.46 | 49.13 | 44.89 | 60.90 | 30.88 | 79.24 | 1.97 | 0.28 | 2.54 |
| | Expt. | -- | -- | -- | -- | -- | -- | -- | -- | -- |
| | Theory [9] | 90.00 | 73.70 | 38.80 | 79.1 | 20.93 | 57.70 | 3.78 | 0.37 | 4.70 |

For cubic crystal, the mechanical stability criteria [30] are given by,

$$C_{11} > 0,\ C_{44} > 0,\ C_{11} - C_{12} > 0 \text{ and } C_{11} + 2C_{12} > 0 \qquad (1)$$

It is clearly seen from Table 3 and Table 4 that our calculated elastic constants satisfy the above restrictions, suggesting that ScM (Rh, Cu and Hg) compounds is mechanically stable up to 60 GPa and ScAg compound is mechanically stable up to 20 GPa. From Table 3, it can be also seen that the elastic constants for ScRh, ScCu, ScAg and ScHg are in reasonable agreement with previous experimental and available theoretical results. In Table 4, we present the calculated elastic constants of ScM compounds under the applied hydrostatic pressure in the range of 0 – 60 GPa with a step of 20 GPa except ScAg. Because ScAg compound becomes unstable under pressure more than 20 GPa. To the best of our knowledge, no experimental or theoretical information on the elastic properties of ScM compounds under pressure for comparison except ScRh is available in the literature. The obtained elastic constants of ScM as the function of the applied hydrostatic pressure are displayed in Fig. 3 and Fig. 4. It can be seen that the elastic constants $C_{11}$, $C_{12}$ and $C_{44}$ almost increases with the applied pressure. The most important elastic parameters of materials such as bulk modulus (B), shear modulus (G), Young's modulus (E), Poisson's ratio (ν) and anisotropy factor (A) are obtained from the calculated elastic constant $C_{ij}$ using the Voigt-Reuss-Hill (VRH) method [31]. For the cubic crystal, the Voigt bounds [32] and Reuss bound [33] of the bulk modulus and shear modulus are given as:

$$B_v = B_R = \frac{(C_{11} + 2C_{12})}{3} \qquad (2)$$

$$G_v = \frac{(C_{11} - C_{12} + 3C_{44})}{5} \qquad (3)$$

The Reuss bounds of the bulk modulus and shear modulus are:

$$B_v = G_v \qquad (4)$$

$$\text{and,} \quad G_R = \frac{5C_{44}(C_{11} - C_{12})}{[4C_{44} + 3(C_{11} - C_{12})]} \qquad (5)$$

The expression of bulk modulus B and shear modulus G are given as follows:

$$B = \frac{1}{2}(B_R + B_v) \tag{6}$$

$$G = \frac{1}{2}(G_v + G_R) \tag{7}$$

Using the bulk modulus B and shear modulus G, the Young's modulus E and Poisson's ratio (ν) and anisotropic factor (A) are obtained according to the following formula [34]

$$E = \frac{9GB}{3B + G} \tag{8}$$

$$\nu = \frac{3B - 2G}{2(3B + G)} \tag{9}$$

$$A = \frac{2C_{44}}{(C_{11} - C_{12})} \tag{8}$$

The computed values of the bulk modulus B, shear modulus G, Young's modulus E, Poisson's ratio ν and anisotropic factor A are presented in Table 4.

**Table 4.** The calculated elastic constants, bulk modulus *B* (GPa), shear modulus *G* (GPa), Young's modulus *E* (GPa), *B/G*, Poisson's ratio *v* and anisotropy factor *A* of ScM (M = Rh, Cu, Ag, Hg) compounds under hydrostatic pressure.

| Materials | P (GPa) | $C_{11}$ | $C_{12}$ | $C_{44}$ | B | G | E | B/G | v | A |
|---|---|---|---|---|---|---|---|---|---|---|
| ScRh | 20 | 325.55 | 162.44 | 78.59 | 216.81 | 79.76 | 213.14 | 2.71 | 0.28 | 0.96 |
|  | 40 | 417.16 | 230.15 | 95.87 | 292.48 | 94.91 | 256.93 | 3.08 | 0.35 | 1.02 |
|  | 60 | 514.98 | 288.82 | 129.77 | 364.20 | 122.81 | 331.20 | 2.96 | 0.34 | 1.14 |
| ScCu | 20 | 198.16 | 145.38 | 84.05 | 162.97 | 52.91 | 143.22 | 3.08 | 0.35 | 3.18 |
|  | 40 | 231.99 | 227.67 | 93.91 | 229.11 | 31.21 | 89.56 | 7.34 | 0.43 | 43.47 |
|  | 60 | 353.81 | 258.67 | 121.69 | 290.00 | 83.50 | 228.56 | 3.47 | 0.36 | 2.55 |
| ScAg | 20 | 157.25 | 55.01 | 32.10 | 153.47 | 21.46 | 61.51 | 7.15 | 0.43 | 20.73 |
|  | 40 | -- | -- | -- | -- | -- | -- | -- | -- | -- |
|  | 60 | -- | -- | -- | -- | -- | -- | -- | -- | -- |
| ScHg | 20 | 180.21 | 136.96 | 84.92 | 151.37 | 49.35 | 133.53 | 3.06 | 0.35 | 3.92 |
|  | 40 | 257.90 | 206.91 | 109.84 | 223.92 | 58.22 | 160.72 | 3.84 | 0.38 | 4.30 |
|  | 60 | 328.85 | 272.72 | 135.44 | 291.43 | 73.00 | 202.12 | 3.99 | 0.39 | 4.82 |

The bulk modulus B can measure the hardness of material [35]. From Table 4, it can be seen that the bulk modulus B increases with the increase in pressure, which predicate that the intermetallic compounds ScM (M = Rh, Cu, Ag and Hg) becomes more difficult to change volume with the increasing pressure. The pressure dependence of the bulk modulus B, shear modulus G and Young's modulus E are displayed in Fig. 3 and Fig.4. As shown in figure the bulk modulus of ScRh is larger meaning that it is harder among ScM compounds. The shear modulus G reflects the resistance to reversible deformations upon shear stress [36]. The obtained bulk modulus and shear modulus behaves with linear relationships with the applied hydrostatic pressure. The Young's modulus is defined as the ratio of the tensile stress to the tensile strain and is used to measure of the stiffness of the materials. The larger the Young's modulus E, the stiffer the material is. The Young's modulus increases with pressure, meaning that the pressure has a significant effect as the stiffness of ScM compounds. The calculated values of G and E for ScAg compound could not be compared due to the lack of experimental and theoretical data.

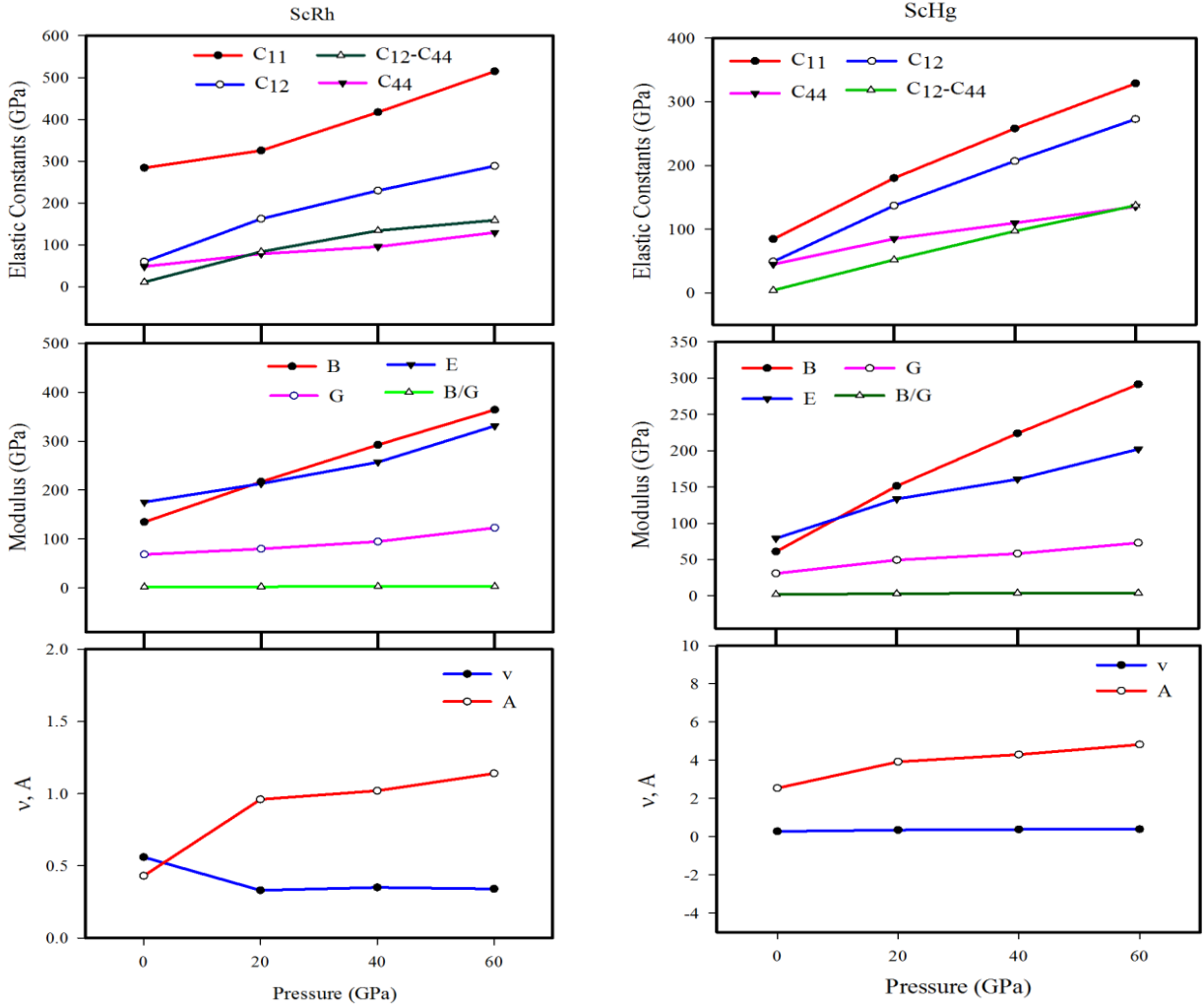

**Fig. 3.** The calculated elastic constants and elastic parameters (B, G, E, ν, A) of ScRh and ScHg compounds under different pressure.

Pugh [37] have suggested that the ratio of bulk modulus to shear modulus is used to predict the ductile and brittle character of materials. The critical value to distinguish brittle and ductile material is 1.75. If the value of B/G < 1.75, the material shows in a brittle character and if B/G > 1.75, it behaves in a ductile manner. From Table 4 it can be seen that, all the values of B/G are larger than the critical value 1.75, which mean that the ScM (M = Rh, Cu, Ag and Hg) compounds behaves in a ductile manner. In our work, the calculated values of the ratio B/G increases of applied pressure, the value increases slowly, indicating that the pressure can improve the ductility and weaken the brittleness of ScM (Rh, Cu, Ag and Hg) compounds.

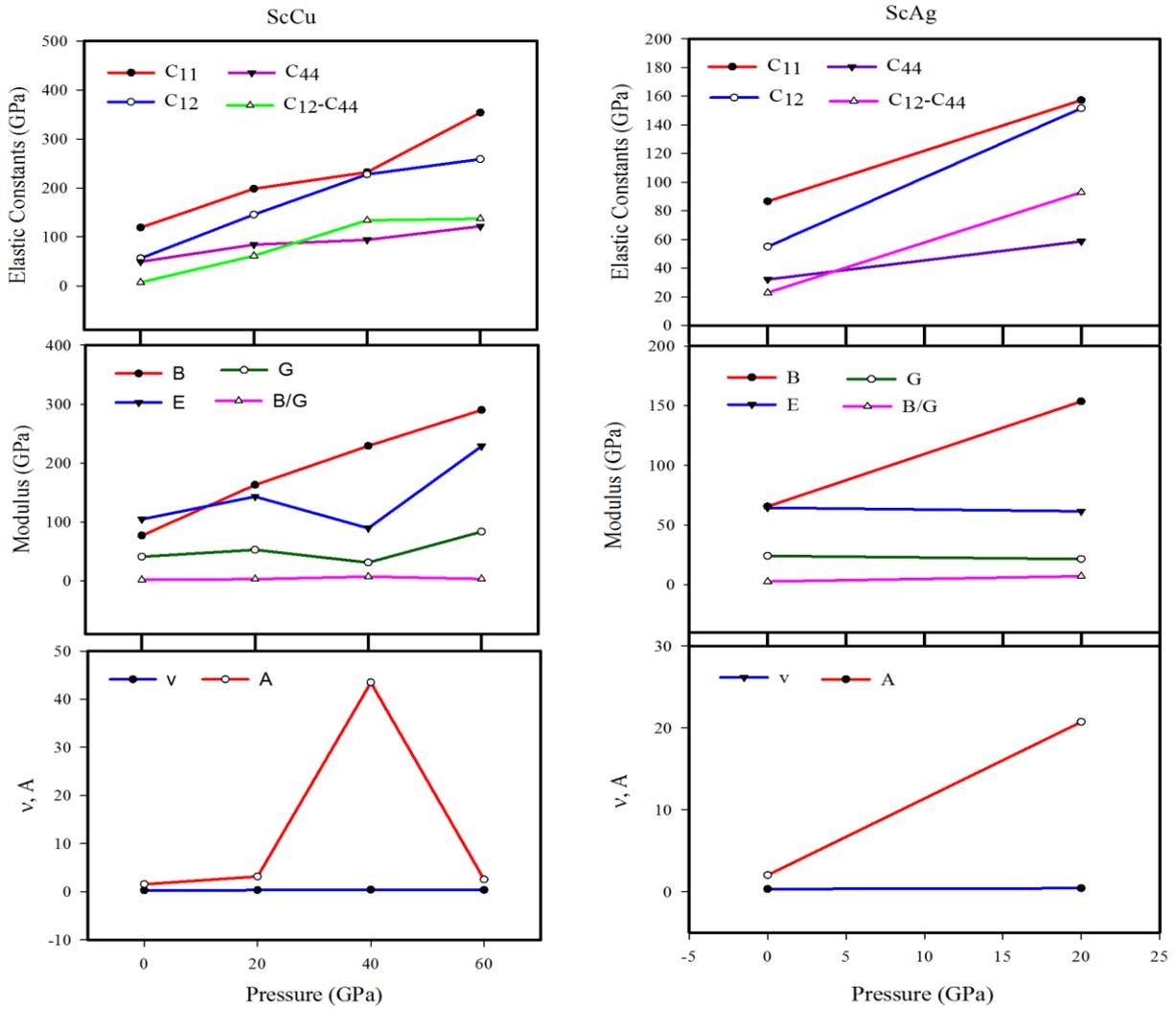

**Fig.4**. The calculated elastic constants and elastic parameters (B, G, E, ν, A) of ScCu and ScAg compounds under different pressure.

Poisson's ratio ν is an important parameter which gives a measure of the stability of the material. J. F. Nye [38] proposed that the larger the Poisson's ratios ν, better the plasticity is. The lower limit of Poisson's ratio ν is 0.25 and the upper limit is 0.50 for central forces in solids [39]. Our calculated values of Poisson's ratios of ScM at zero pressure are larger than the lower limit, which mean that the interatomic forces of ScM are central forces. At pressure increases, the Poisson's ratios ν will larger than lower limit and less than upper limit indicates that interatomic forces are central at high pressure. In addition, the elastic anisotropy A is an important factor affecting the mechanical stability of materials [40]. The anisotropy factor is an important implication in engineering science as well as material science. The anisotropy factor is used to measure of the degree of anisotropy in materials [41]. When the value of A = 1 means a completely isotropic material. From Table 4, it can be seen that the values of A is less than 1 at 0 GPa and 20 GPa for ScRh compound indicating that, the ScRh compound is elastically anisotropic material. From 20 GPa to 60 GPa for ScRh compound and other (ScCu, ScAg and ScHg) compounds from 0 GPa to 60 GPa, the values of A is larger than 1 (see table 4), so we conclude that the compounds are strongly anisotropic.

## 3.3. Optical properties

The details study of the optical properties is crucial for better understanding of the electronic structure of materials. In this section we have done a comparative comprehensive study about various optical constants of ScRh, ScCu, ScAg and ScHg respectively. The optical properties of a compound can be obtained from the complex dielectric function *ε (ω)* which is given by *ε (ω) = ε$_1$ (ω) + iε$_2$ (ω)*, where *ε$_1$ (ω)* is the real part and *ε$_2$ (ω)* is the imaginary part. *ε$_2$ (ω)* is obtained from the momentum matrix elements between the occupied and the unoccupied electronic states and calculated directly using [42]

$$\varepsilon_2(\omega) = \frac{2e^2\pi}{\Omega\varepsilon_0} \sum_{k,v,c} |\psi_k^c| u.r |\psi_k^v|^2 \delta(E_k^c - E_k^v - E) \qquad (9)$$

Where, *u* is defined as the polarization vector of the incident electric field, *ω* is the frequency of light, *e* is defined as the electronic charge and $\psi_k^c$ and $\psi_k^v$ are the conduction and valence band wave functions at *k*, respectively. The optical constants such as refractive index, loss-function, absorption spectrum, conductivity and reflectivity are obtained by using Eqs. 49 to 54 in ref. [42]. The optical functions of those compounds under study are calculated for photon energies up to 80 eV to the direction of polarization vector [100]. 0.5 eV gaussain smearing was used for all calculations.

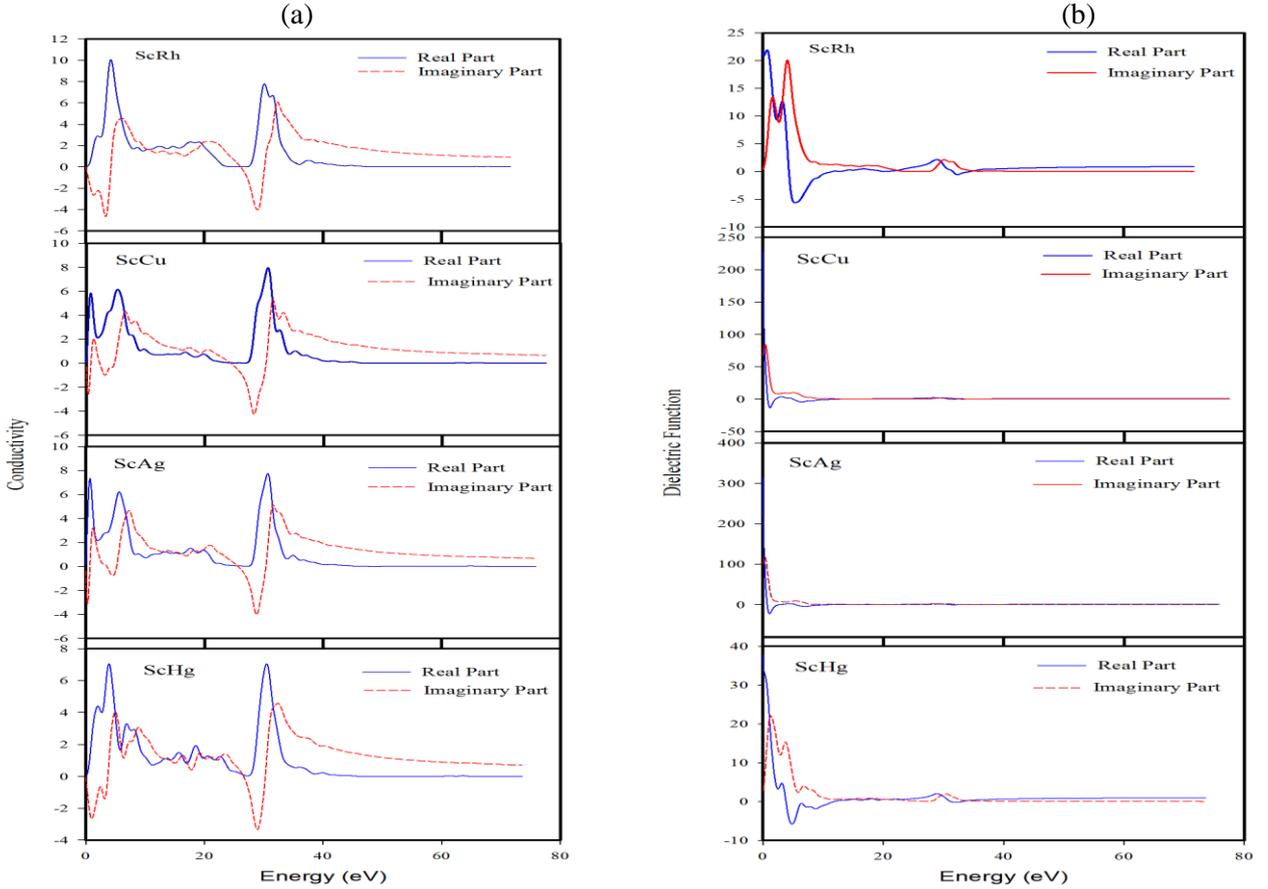

**Fig. 5.** The conductivity (a), and dielectric function (b) of ScM (M = Rh, Cu, Ag, Hg) for polarization vector [100].

Fig. 5(a) represents the conductivity spectra of ScM (M = Rh, Cu, Ag, Hg) as a function of photon energy. From the plot it is evident that the photoconductivity for all of the four compounds starts with zero photon energy indicating the compounds under investigation have zero band gaps. Several maxima and minima are observed in the conductivity plot. We observe similar peak in the conductivity plot for all of the four compounds within the energy range studied. However, electrical conductivity of these materials increases as a result of absorbing photons [43].

The dielectric function is a very crucial optical constant which explains what an electric field does to a material [44]. The real part of the dielectric function describes about the polarization of material due the applied electric field. The imaginary part is related to the absorption in a material. Fig. 5(b) shows the dielectric function of ScM (M = Rh, Cu, Ag, Hg). It is observed from the figure that the value of $\varepsilon_2$ becomes zero at about 11 eV, 2eV, 1.98 eV, and 11 eV for ScRh, ScCu, ScAg and ScHg respectively. These values indicate that these materials become transparent above these certain values. When the value of $\varepsilon_2(\omega)$ becomes nonzero, then absorption begins. From Fig 5(b) it is evident that all of the four compounds show slight absorption within the energy range studied. The values of static dielectric constant are 21, 225,315 and 38 for ScRh, ScCu, ScAg and ScHg respectively. Materials with high dielectric constants are useful for manufacturing of high value capacitors [44]. In this case the priority is ScAg > ScCu > ScHg > ScRh.

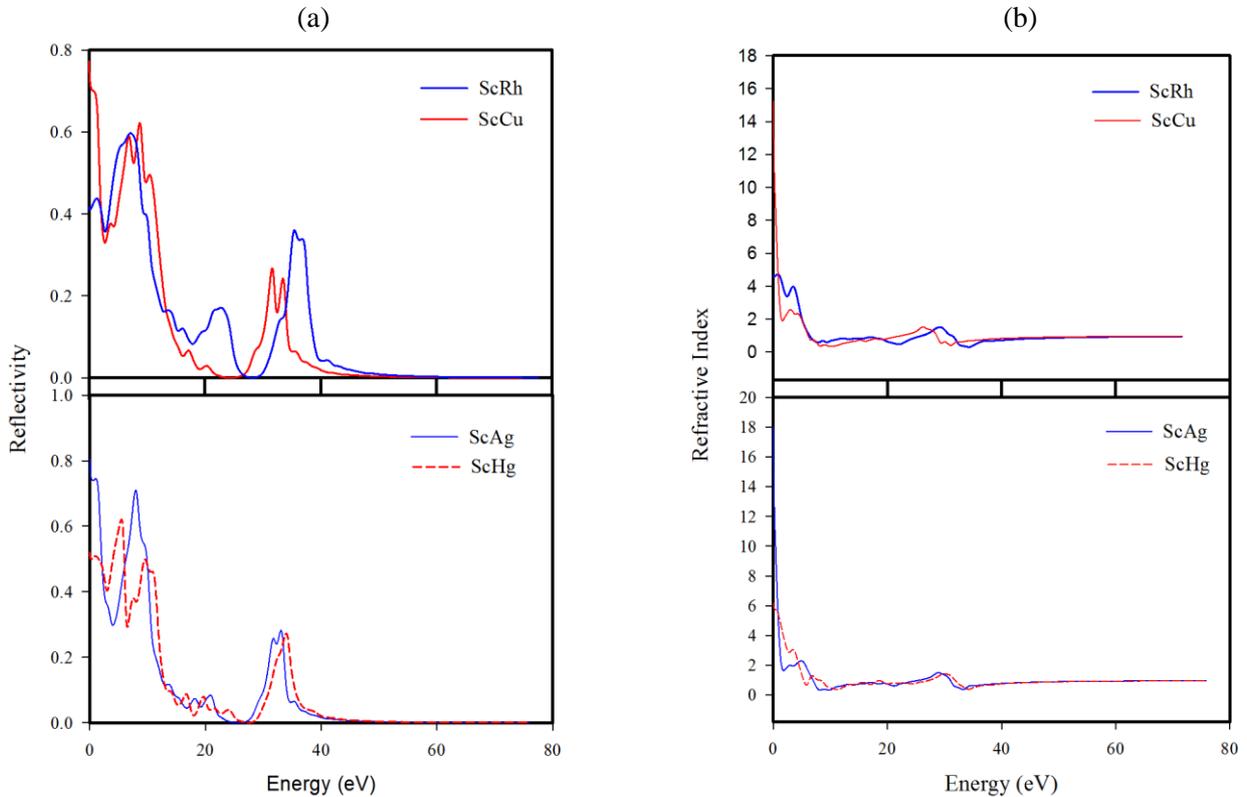

**Fig. 6.** The Reflectivity (a), and Refractive index (b) of ScM (M = Rh, Cu, Ag, Hg) for polarization vector [100].

Fig. 6(a) shows the reflectivity spectra as a function of photon energy of ScM (M = Rh, Cu, Ag, Hg). We see that the reflectivity is 0.41-0.43 in the infrared region for ScRh, 0.27-0.54 for ScCu, 0.83-0.62 for ScAg and 0.50-0.48 for ScHg. These values drop rapidly in the high energy region with some peaks as a result of intraband transition.

The refractive index is an important optical function of a material which determines how much light is bent, or refracted, when entering a material. Fig. 6(b) illustrates the refractive index of ScM (M = Rh, Cu, Ag, Hg). Form figure we obtain the values of static refractive index as 4.75, 15, 18, and 6 for ScRh, ScCu, ScAg and ScHg respectively. The refractive index of all these compounds is higher in the low energy region and gradually decreased in the high energy region.

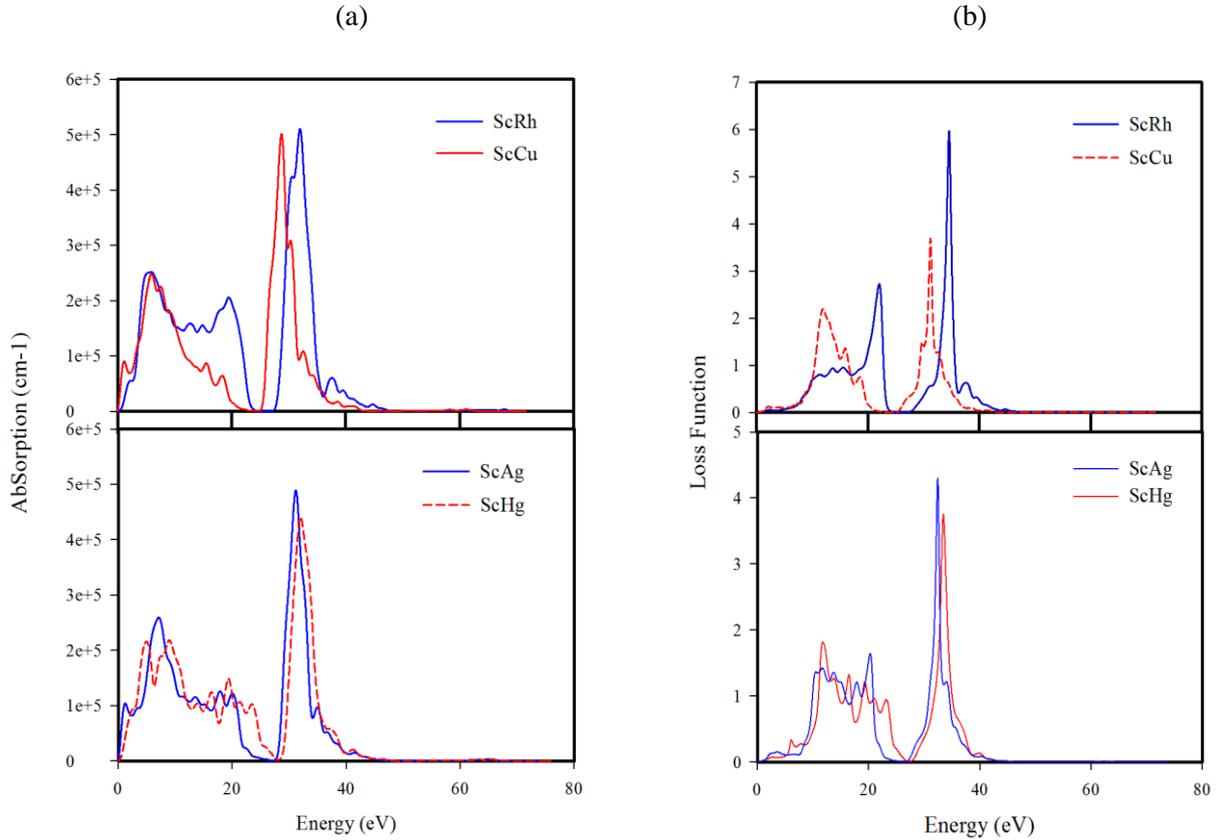

**Fig. 7.** The Absorption coefficient (a), and Loss function (b) of ScM (M = Rh, Cu, Ag, Hg) for polarization vector [100].

Fig. 7(a) illustrates the absorption coefficient spectra of ScM (M = Rh, Cu, Ag, Hg) as a function of photon energy. It provides data about optimum solar energy conversion efficiency and indicates how far light of a specific energy (frequency) can penetrate into the material before being absorbed [45]. The absorption spectra indicate that all of the compounds under study exhibits metallic nature since the spectra starts from 0 eV for all of the compounds. Several peaks are observed in the plot but the highest peaks are located at 32, 29, 30.88, and 3.62 eV for ScRh, ScCu, ScAg and ScHg respectively. All of these compounds possess good absorption coefficient in the high energy region.

The loss function is an important optical parameter which explains the energy loss of a fast electron traversing a material. It is shown in Fig. 7(b) as a function of photon energy. The highest peaks are found at 34, 31, 32, and 33 eV for ScRh, ScCu, ScAg and ScHg respectively. These peaks are related to the feature that is associated with plasma resonance, and the corresponding frequency is called bulk plasma frequency [46].

## 4. Conclusion

In this work, first principles calculations have been used to investigate of the influence of pressure on the structural, elastic and optical properties of ScM (M = Rh, Cu, Ag and Hg) compounds. The structural properties including lattice parameter, cell volume, bulk modulus and first order pressure derivative of the bulk modulus were calculated successfully. The calculated lattice parameters are in good consistent with the experimental and available theoretical values. The pressure dependence of lattice constants and cell volume are also investigated and all of them decrease with the increasing pressure.

Our calculated results of elastic constants shows that the structure of intermetallic ScM (M = Rh, Cu, and Hg) and ScAg compounds are mechanically stable with the pressure range from 0 to 60 GPa and 0 to 20 GPa respectively. It is worthy to mention that the ScAg material becomes unstable under the pressure more than 20 GPa. The pressure effects on elastic modulus (B, G, and Y), B/G, Poisson's ratio $v$ and elastic anisotropy $A$ have also been investigated. The ductility of these ScM compounds have been analyzed using the Pugh's rule; the results shows that all the materials are ductile in nature.

The study of optical properties reveals that all of these compounds possess good absorption coefficient in the high energy region and the refractive index of all these compounds is higher in the low energy region and gradually decreased in the high energy region. The study of conductivity reveals that the photoconductivity for all of the four compounds starts with zero photon energy indicating the compounds under investigation have zero band gaps.